\documentclass[twocolumn,aps]{revtex4}
\usepackage{graphicx}

\begin{document}

\title{Noise-induced Transition between Dynamic Attractors in the
Parametrically Excited Magneto-optical Trap}

\author{Kihwan Kim, Myungsun Heo, Ki-Hwan Lee, Hyoun-Jee Ha,
Kiyoub Jang, Heung-Ryoul Noh$^{\dag}$,
and Wonho Jhe\footnote{Corresponding author: whjhe@snu.ac.kr}}

\affiliation{School of Physics and Center for Near-field
Atom-photon Technology,\\
Seoul National University, Seoul 151-747, Korea \\
$^{\dag}$Department of Physics, Chonnam National University,
Gwangju 500-757, Korea}

\begin{abstract}
We have investigated noise-induced transition of atoms between
double or triple phase-space attractors that are produced in the
parametrically driven magneto-optical trap. The transition rates
between two or three dynamic attractors, measured for various
modulation frequencies and amplitudes, are in good agreement with
theoretical calculations and Monte-Carlo simulations based on the
Langevin equations. Our experiment may be useful to study
nonlinear dynamic problems involving transitions between states
far from equilibrium.

\end{abstract}

\date{\today}

\maketitle

There have been much studies on fluctuation-induced transition
between states that are in equilibrium or far from equilibrium.
For equilibrium systems, after Kramers' seminal work
\cite{Kramers}, many theories have been suggested \cite{Haenggi}
and tested in many different experiments: for example, optically
trapped Brownian particles \cite{Simon, McCann}, analogue
circuits \cite{Dykman}, and semiconductor lasers \cite{Hales}.
Recently, there have been several experiments on far from
equilibrium systems, such as Penning trap \cite{Penning},
vibration-fluidized granular matter \cite{Norl}, and Josephson
junction \cite{Josephson}. Theoretical works on calculation of the
transition rates \cite{Dykman2} or the transition paths
\cite{Freidlin}, and analysis of the critical exponents
\cite{Dykman3} have been performed. In particular, interesting
phenomena including oscillatory behavior with respect to the
noise intensity \cite{Stein} or the phase \cite{Golding}, and
saddle-point avoidance \cite{Luchinsky} have been also expected.
However, except the Penning trap, the only quantitatively
investigated experimental system was the analogue electrical
circuit \cite{Luchinsky2}, which can be considered as analogue
simulations. For the Penning trap, on the other hand, experiment
was carried out within a very narrow parameter region: only near
the bifurcation points.

In this Letter, we report on the experimental study of the
noise-induced transition in the paramterically driven
magneto-optical trap (MOT) which is a far from equilibrium system.
In particular, we have investigated the transition in nearly full
parameter regions from super-critical bifurcation points (dynamic
double attractors) to sub-critical bifurcation points (dynamic
triple attractors). For double attractors, we have measured the
transition rates by observing directly the change of population
difference between two oscillating wells at various modulation
frequencies and modulation amplitudes. On the other hand, the
transition between three attractors, which has never been probed
experimentally to our knowledge, has been studied indirectly by
measuring the populations in each well \cite{Dykman2}. We have
also investigated the transition during one period of modulation
and found no phase dependence of the transition rates
\cite{Golding}. All the experimental results are found in good
agreement with theoretical calculations and Monte-Carlo
simulations based on the Langevin equations, which describe well
the quantum dissipation \cite{DCohen} including spontaneous
emission that is the origin of fluctuations of atomic motion in
MOT.

As reported in previous papers \cite{cnat}, when the
intensity-modulation frequency $f$ of the cooling laser is around
twice the trap frequency between $f_1$ and $f_2$, atoms are
separated into two clouds [Fig. 1(a) Left] and oscillate in
out-of-phase motion, which corresponds to the limit-cycle (LC)
characterized by dynamic double attractors. Due to the
nonlinearity of MOT, as $f$ is increased above $f_2$ but below
$f_3$, there appears an additional stable attractor at the center
of the LC motion, called sub-critical bifurcation [Fig. 1(a)
Right]. Given the modulation amplitude $h$, the corresponding
frequencies are, $f_{1}$ = $2f_{0}-(f_{0}/2)\sqrt{h^{2}-h_{T}^2}$,
$f_{2}$ = $2f_{0} +(f_{0}/2)\sqrt{h^{2}-h_{T}^2}$, and $f_{3}$ =
$f_{0}+(h f_0 /2 h_{T})\sqrt{4+h_{T}^{2}}$, where $f_0$ is the
natural trap frequency of MOT, and
$h_{T}= \beta /(\pi f_{0})$ is the threshold value of $h$ for
parametric resonance to occur for a given damping coefficient
$\beta$.

Assuming no fluctuations of atomic motion, initial conditions of
atomic position and velocity determine which attractors
(represented by red dots) an atom ends up with. Figure 1(b) shows
each region of two attractors (Left) and that of three attractors
(Right). For example, if the initial condition of an atom lies in
the light gray region, the atom approaches the attractor located
in the same basin. In reality, however, there exist atomic
fluctuating motions due to spontaneous emission, resulting in
broadened distributions of atomic position and velocity near the
stable attractors. For large diffusion (or spontaneous emissions),
certain atoms may jump far from the original attractor and be
transferred to another attractor through the unstable regions near
the boundary. In this case, the shape of the atomic phase-space
distribution resembles a dumbbell which has a narrow neck at the
unstable point. The distribution represents the dynamic potential
wells, as discussed in the static potential experiment
\cite{McCann}.

The transition rate $W_n$ can be characterized by the activation
law, $W_n \propto \exp(-S_n/D)$, which is the dynamic version of
the Kramers' equation. Here $S_n$ is the activation energy of each
state $n$ and $D$ is the amount of diffusion which is proportional
to the diffusion constant of MOT.
In what follows we set $n=1$
and 2 for the stable states of two dynamic attractors and $n=0$
for the stationary state at the MOT center. As observed in the
stationary position-space potentials, the transition rate
increases as the diffusion increases or the activation energy
decreases. Note that since all the nonlinear terms in the
equations of motion of MOT have the same sign as the first order
term, the activation energy is proportional to $f$ and $h$.

Unlike the single electron in the Penning trap, there are more
than 10$^7$ atoms in the initial MOT and each cloud of the LC
motion under parametric excitation is almost equally populated.
In order to monitor the transition between two dynamic
attractors, it is needed to blow away one of the clouds because
the atomic number transferred from attractor 1 to 2 is the same
as that from 2 to 1.
When atoms, say in attractor 1, are removed,
one can observe some atoms in attractor 2 are transferred to 1
with time. Due to the two-way transitions the population
difference decreases exponentially. For many particle systems the
transitions can be described by the following simple rate
equations,
\begin{eqnarray}
\frac{dN_{1}}{dt}&=&R-\gamma N_{1} -W_{1} N_{1} +W_{2} N_{2} \, ,
\nonumber \\
\frac{dN_{2}}{dt}&=&R-\gamma N_{2} -W_{2} N_{2}+W_{1} N_{1}\, ,
\label{rateeq}
\end{eqnarray}
where $N_1$ ($N_2$) is the population of the attractor 1 (2), $R$
is the trap loading rate, $\gamma$ is the loss rate due to
collisions by the background atoms, and $W_{1}$ ($W_{2}$) is the
transition rate from the attractor 1 to 2 (2 to 1). Since $R$ and
$\gamma$ are small in our experiment (as discussed later), we
neglect these terms and also assume $W = W_{1} = W_{2}$ due to the
symmetry. Then the steady-state solution of the population
difference $\Delta N$(= $N_2-N_1$) is given by
\begin{eqnarray}
\Delta N &=& \Delta N_{0} e^{-2 W t} \, , \label{solution}
\end{eqnarray}
where $\Delta N_{0}$ is the difference with one attractor (say,
$n=1$ state) empty, or $N_1$ = 0.


\begin{figure}[ht]
\centering
\includegraphics[scale=0.4]{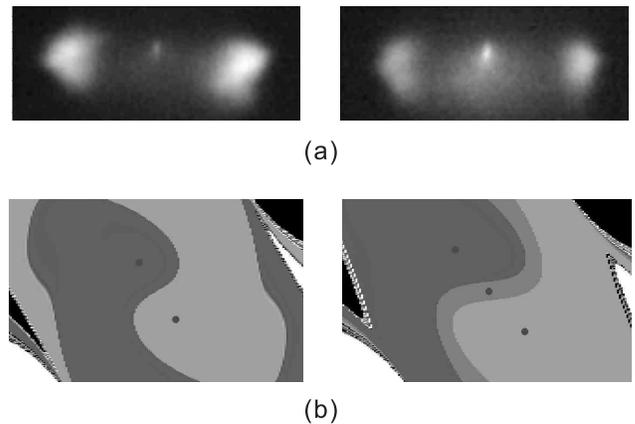}
\caption{{(a) The atomic fluorescent images of the LC motion when
there are two attractors (Left) and three attractors (Right). (b)
The phase-space map of two (Left) and three attractor (Right)
conditions. }} \label{fig:1initialMAP}
\end{figure}

Our experimental setup is similar to that described in previous
work \cite{cnat}, but with new features: we have used a photodiode
array \cite{TF} to measure the transition rates and used a
resonant laser to blow away selectively one atomic cloud (say,
$n=1$). The blowing laser was cylindrically focused at 5 mm left
from the center of the LC motion. The intensity of the laser was
over 5 times saturation intensity and it was turned on for 3 ms
in order to remove only one atomic cloud.

\begin{figure}[ht]
\centering
\includegraphics[scale=0.4]{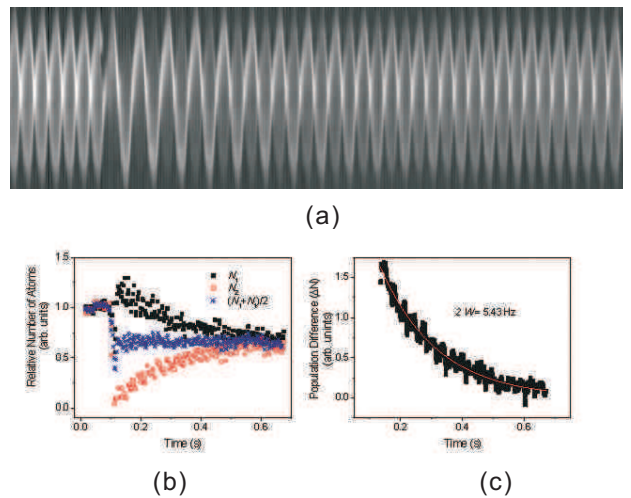}
\caption{{(a) Typical contour plot of absorption signals obtained
at $f$ = 63 Hz ($f/f_{0}$ = 1.9) and $h$ = 0.7. (b) Decay of the
population of state 2 (filled box), growth of the atomic number of
the empty state 1 (empty box) as measured from (a), and the
averaged half of the total number (asterisk). (c) The population
difference between the two states shows exponential decay. }}
\label{fig:2TypicalResult}
\end{figure}

Experimental results for double (triple) attractors are obtained
at the intensity of $0.034$ $I_s$ ($0.06$ $I_s$) and the detuning
of -2.7 $\Gamma$ for the cooling laser along the longitudinal
$z$-axis, and the magnetic field gradient of 10 G/cm, where $I_s$
= 3.78 mW/cm$^2$ is the averaged saturation intensity and
$\Gamma$ = $2 \pi \times$ 5.9 MHz is the natural line width. The
laser intensity and detuning along the transverse $x$- and
$y$-axis are 0.63 $I_s$ and -3.0 $\Gamma$, respectively. For
these parameters, the measured trap frequency is 33.4 Hz (43.9
Hz, for triple states), and the detected damping coefficient is
45.4 s$^{-1}$ (for triple states, 85.9 s$^{-1}$), which are in
good agreement with the simple Doppler theory. Note that the
sub-Doppler nature of MOT is dramatically suppressed and thus
neglected when the transverse laser detuning is slightly different
from that of the $z$-axis laser, as discussed in depth in our
previous work \cite{TF}.

The typical data about the transition between dynamic double
wells are presented in Fig. 2. Figure 2(a) shows a contour plot
generated by the atomic absorption signals recorded on the 16
channel photo-diode array. The vertical axis represents the atomic
position and the longitudinal axis the time evolution. In Fig.
2(a), large absorption (or large atomic number) is indicated by
bright color. From the plot one can trace the oscillating LC
attractors and measure the atomic number. It can be observed that
one attractor is made empty at 7 half cycles from the left, which
is then repopulated with time, as plotted in Fig. 2(b). Figure
2(c) presents the temporal variation of the population difference
($\Delta N$), which clearly shows the exponential decay as
expected in Eq. (\ref{solution}) where the decay rate is twice
the transition rate. Note that the total number of atoms is
nearly conserved during the transitions [asterisks in Fig. 2(b)],
which justifies neglect of the loading rate $R$ and loss rate
$\gamma$ in Eq. (\ref{rateeq}). In reality, the loss rate of our
MOT is 0.15 s$^{-1}$ which is much smaller than the typical
transition rates.

\begin{figure}[ht]
\centering
\includegraphics[scale=0.4]{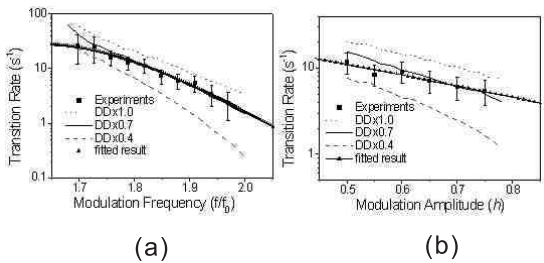}
\caption{{ Transition rates for various (a) modulation frequencies
and (b) modulation amplitudes, compared with the Monte-Carlo
simulations for scaled diffusion $D$. }} \label{fig:3MainResult}
\end{figure}

The experimental results in Fig. 3 were obtained by exponential
fitting of Fig. 2(c) for (a) various modulation frequencies at
$h=0.72$ and (b) modulation amplitudes at $f/f_{0}=1.89$. The
curves in Fig. 3 represent the Monte-Carlo simulation results
obtained from simple Doppler equations based on two-level atom and
random spontaneous emission. We have first calculated the
trajectories of atomic motion of $10^4$ atoms initially in $n=2$
state and monitored the subsequent time evolution of the
population of the two states. We have then obtained the transition
rates by fitting Eq. (\ref{solution}). As can be observed in Fig.
3, the experimental results are in good agreement with the
simulations when diffusion constant is  0.7 times DD, where DD
represents the Doppler diffusion coefficient (Eq. (22) in Ref.
\cite{OC04}).



In the low damping regime, one can also find that the transition
rates approximately vary around the bifurcation point as
\begin{equation}
W = c_1 \exp \left[ - \frac{c_{2}}{hD} (f- f_1 )^2 \right] \, ,
\end{equation}
where $f_1$ is the modulation frequency at the bifurcation point,
and  $c_1$ and $c_2$ are constants \cite{Dykman2}. The constant
$c_2 = \left(1/f_0^2 \right) \sqrt{2/h_{T}}$ was obtained by
direct calculation \cite{Dykman2}. The constant $c_1$ and the
diffusion $D$ were used as fitting parameters. The consequent
diffusion $D$ for our experimental conditions is 0.071(8), which
are in the same order to the calculated value with simple Doppler
theory $D= 3A_{0}/\left(2m^2 \omega_0 \beta^2 \right) {\rm DD}$
\cite{Dykman2}. Here the coefficient $A_0$ is nonlinear
coefficient of a term $z^3$ in the equation of motion of a
parametrically-driven atom, which is presented as Eq. (7) in Ref.
\cite{OC04}. The triangles in Fig. 3(a) and 3(b) represent the
fitted results. Even though the calculations are mainly well
applied in the limit $f \gtrsim f_1 $ and $\beta \ll \omega_0$, we
can see these fittings very well describe the experimental
results.


Since $W$ and $f$ are of the same order in our experiment, we also
have studied the time dependence of the transition during one
period of parametric oscillation. By averaging the absorption
signals over 20 times, we have obtained the population evolution
during each period. The experiments were performed at $f = 1.76
f_{0}$ and $h=0.72$, with the same laser conditions and magnetic
field gradient as above. The transition rate $W$ was 15.4
s$^{-1}$, which was measured during three oscillation periods
until the population difference became negligible. In particular,
we have not found any phase dependence of the transition rates,
which indicates that transitions occur constantly without any
dependence of location of atomic clouds in phase space. Note that
these results are very different from those observed in periodic
driven systems \cite{Golding}, which were studied only
theoretically. Experimental results are also in very good
agreement with simulations, which do not show any clear variation
of the transition rates during one period.

\begin{figure}[ht]
\centering
\includegraphics[scale=0.4]{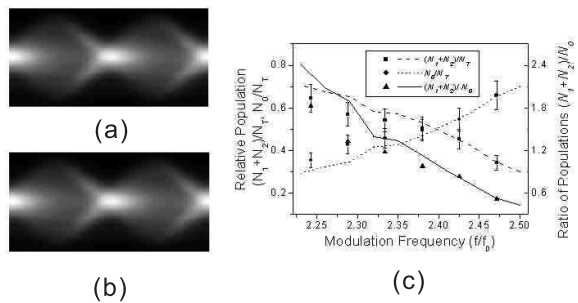}
\caption{{(a) Contour plot of the absorption signals for triple well
conditions, where $f$=106 Hz (= 2.42 $f_{0}$) and $h$=0.72. (b)
Contour plot of the simulation results, where $f$=105.3 Hz (= 2.41
$f_{0}$) and $h$=0.7. (c) Relative populations of state 1 and 2
($N_1$ + $N_2$)/$N_T$ compared to that of state 0 ($N_0$/$N_T$),
where points (filled box and filled circle) represent the
experimental results and curves the simulations. Filled triangles
show the experimental ratios of $N_1$ + $N_2$ to $N_0$ and curves
the simulation results. }} \label{fig:4TripleWells}
\end{figure}

For the case of dynamic triple attractors, direct observation of
the transition is difficult in our setup. When $f_2$ $<$ $f$ $<$
$f_3$, it is numerically observed that transition occurs only
between one of the two dynamic states (1 or 2) and the stable
state (0), with no direct transitions from 1 (2) to 2 (1).
Moreover, unlike the double wells, the transition rates $W_{i0}$
from state $i$ to 0 are not necessarily the same as $W_{0i}$ from
state 0 to $i$ ($i=1,2$). Thus the method of just removing the
state 1 and monitoring the time evolution of the population
difference does not provide the necessary information of the
relevant transition rates. Therefore we have studied the
transition in the sub-critical region by using an indirect method
of measuring the populations in each state. The ratios of the
population $N_{1}$ (or $N_{2}$) to $N_{0}$ can be used to obtain
the transition rates, that is, $N_{1}/N_{0}$ = $N_{2}/N_{0}$
=$W_{0}/2W$, where we have employed $W_{10}=W_{20}=W$ and
$W_{01}=W_{02}=W_{0}/2$. Note that for a larger modulation
frequency, $N_{0}$ becomes larger than $N_{1}$ (or $N_{2}$)
\cite{Dykman2}, which is confirmed in our experiment.



The experimental and simulation results for dynamic triple
attractors are presented in Fig. 4. Figure 4(a) shows the motions
of atomic clouds during single period of oscillation, which
provides the population in each state. States 1 and 2 show
out-of-phase oscillating motions, while state 0 stays at the
center. Our simulations [Fig. 4(b)] have very similar results to
the absorption image [Fig. 4(a)], with different spatial
resolutions (1 mm for Fig. 4(a) and 0.1 mm for Fig. 4(b)).

The data of ($N_1$ + $N_2$)/$N_T$ (filled boxes) and $N_0$/$N_T$
(filled circles) in Fig. 4(c) are obtained by fitting the profiles
of Fig. 4(a) at 0.58 $\pi$ phase. The dashed curves (($N_1$ +
$N_2$)/$N_T$) and dotted curves ($N_0$/$N_T$) are derived from
Fig. 4(b) at 0.5 $\pi$. Here $N_{n}$ ($n$ = 0,1,2) is the
population of the state $n$ and $N_{T}$=$N_{0}$+$N_{1}$+$N_{2}$.
The profile shows three peaks, representing three states, are
fitted by triple Gaussian functions. Filled triangles
(experiment) and solid curves (simulation) are just the ratios of
$N_1$ + $N_2$ to $N_0$, which also provides the ratio of the
transitions rates ($W_{0}/W$). One can observe $N_{0}$ is
increased as $f$ increases, whereas $N_1$ and $N_2$ are
decreased. It shows that the transition from state 1 (2) to state
0 becomes easier than that from 0 to 1 (2) as $f$ increases. These
results are in good agreement with theoretical calculation
$W_{0}/W = \exp[(S_{1}-S_{0})/D]$ where $S_{n}$ is the activation
energy of state $n$ \cite{Dykman2}.

The parametrically driven MOT is an ideal experimental realization
of dynamic double and triple attractors under the conditions far
from equilibrium. We have investigated the noise-induced
transition at various $f$ and $h$ from the super-critical to
sub-critical bifurcation region. The results are well described by
the Monte-Carlo simulations based on the Langevin equations and
the simple Doppler theory.
Note that our system may be important and useful to study
nonlinear problems associated with transitions in far from
equilibrium systems. One may also extend our system to the study
of transitions between dynamic states due to non-symmetric
oscillations that occur in more nonlinear situations
\cite{Silchenko}.

\acknowledgements

This work was supported by the Creative Research Initiative of
the Korea Ministry of Science and Technology. The research of H.
R. Noh was financially supported by research fund of Chonnam
National University in 2004.


\begin{references}

\bibitem{Kramers} H. A. Kramers, Physica (Utrecht) {\bf 7}, 284
(1940).

\bibitem{Haenggi} P. H$\textrm{\aa}$nggi, P. Talkner, and M.
Borkovac, Rev. Mod. Phys. {\bf 62}, 251 (1990).

\bibitem{Simon} A. Simon and A. Libchaber, Phys. Rev. Lett. {\bf 68}, 3375
(1992).

\bibitem{McCann} L. I. McCann, M. I. Dykman, and B. Golding, Nature (London)
{\bf 402}, 785 (1999).

\bibitem{Dykman} M. I. Dykman, P. V. E. McClintock, V. N.
Smelyanskiy, N. D. Stein, and N. G. Stocks, Phys. Rev. Lett {\bf
68}, 2718 (1992).

\bibitem{Hales}J. Hales, A. Zhukov, R. Roy, and M. I. Dykman, Phys. Rev. Lett. {\bf 85},
81 (2000).

\bibitem{Penning} L. J. Lapidus, D. Enzer, and G. Gabrielse, Phys. Rev. Lett. {\bf 83},
899 (1999).

\bibitem{Norl} G. D'Anna, P. Mayor, A. Barrat, V. Loreto and F.
Norl, Nature (London) {\bf 424} 909 (2003).

\bibitem{Josephson} N. Gr{\o}nbech-Jensen, M. G. Castellano, {\it et.
al.} Phys. Rev. Lett {\bf 93}, 107002 (2004); 
I. Siddiqi. {\it et. al.}, cond-mat 0312553 (2003).

\bibitem{Dykman2} M. I. Dykman and M. A. Krivoglaz, Sov. Phys. JETP {\bf 50},
30 (1979); M. I. Dykman, C. M. Maloney, V. N. Smelyanskiy, and M.
Silverstein, Phys. Rev. E {\bf 57}, 5202 (1998).

\bibitem{Freidlin} A. D. Ventsel¡¯ and M. I. Freidlin, Usp. Mat. Nauk. {\bf 25}, 5 (1970)
[Russ. Math. Surv. {\bf 25}, 1 (1970)].


\bibitem{Dykman3} M. I. Dykman, B. Golding, and D. Ryvkine, Phys.
Rev. Lett. {\bf 92}, 080602 (2004).

\bibitem{Stein} R. S. Maier and D. L. Stein, Phys. Rev. Lett. {\bf 77}, 4860 (1996).

\bibitem{Golding} V. N. Smelyanskiy, M. I. Dykman, and B. Golding, Phys. Rev. Lett. {\bf 82}, 3193 (1999).

\bibitem{Luchinsky} D. G. Luchinsky, {\it et. al.}, Phys. Rev. Lett.
{\bf 82}, 1806 (1999).

\bibitem{Luchinsky2} D. B. Luchinsky, and P. V. E. McClintock, Nature
(London) {\bf 389}, 463 (1997); D. B. Luchinsky, P. V. E.
McClintock, and M. I. Dykman, Rep. Prog. Phys. {\bf 51}, 889
(1998).

\bibitem{DCohen} D. Cohen, Phys. Rev. Lett. {\bf 78}, 2878 (1997).

\bibitem{cnat} K. Kim, H. -R. Noh, Y. -H. Yeon, and W. Jhe, Phys. Rev. A {\bf 68}, 031403(R) (2003); K.
Kim, H. -R. Noh, H. J. Ha, and W. Jhe, Phys. Rev. A {\bf 69}, 033406
(2004).

\bibitem{TF} K. Kim, H. -R. Noh, and W. Jhe, to be published in Phys. Rev. A (2005); K.
Kim, K. -H. Lee, M. Heo, H. -R. Noh and W. Jhe, submitted.

\bibitem{OC04} K. Kim, H. R. Noh, and W. Jhe, Opt. Commun. {\bf 236}, 349 (2004).

\bibitem{Silchenko} A. N. Silchenko, S. Beri, D.G. Luchinsky, and P.V. E.
McClintock, Phys. Rev. Lett. {\bf 91}, 174104 (2003).
\end{references}
\end{document}